\documentclass{PoS}

\usepackage{amsmath,multirow}

 \usepackage{amssymb, fontenc, times, mathptmx, graphicx}

\usepackage{amssymb, amsmath}
\usepackage{graphics,epsfig,ulem}
\usepackage{float}

\usepackage{multirow}
\usepackage{longtable}
\usepackage{color}
\usepackage{xcolor}
\usepackage{slashed}

\title{Setting the Stage for a Non-Supersymmetric UV-Complete String
Phenomenology}

\ShortTitle{Stable Non-SUSY Strings}

\author{\speaker{Steven Abel}\\
IPPP, Durham University, Durham, DH1 3LE, UK\\
        E-mail: \email{s.a.abel@durham.ac.uk}}
        
\author{Keith R. Dienes\\
        Department of Physics, University of Arizona, Tucson, AZ  85721  USA \\
        Department of Physics, University of Maryland, College Park, MD 20742 USA\\
        E-mail: \email{dienes@email.arizona.edu}}

\author{Eirini Mavroudi\\
        IPPP, Durham University, Durham, DH1 3LE, UK\\
        E-mail: \email{irene.mavroudi@durham.ac.uk}}

\abstract{
In this talk, I discuss our recent work concerning the construction of non-supersymmetric heterotic string models which have exponentially suppressed dilaton tadpoles and cosmological constants, and thus greatly enhanced stability properties.   The existence of such models opens the door to non-supersymmetric string model-building, and I discuss how semi-realistic string models resembling the Standard Model or any of its unified variants may be constructed within this framework.   These models maintain modular invariance and exhibit a misaligned supersymmetry which ensures UV finiteness, even without spacetime supersymmetry.   I also discuss the potential implications for phenomenology.}

\FullConference{18th International Conference From the Planck Scale to the Electroweak Scale \\
		25-29 May 2015\\
		Ioannina, Greece }

\begin{document}

\def\beq{\begin{equation}}
\def\eeq{\end{equation}}
\def\beqn{\begin{eqnarray}}
\def\eeqn{\end{eqnarray}}
\def\half{{\textstyle{1\over 2}}}
\def\quarter{{\textstyle{1\over 4}}}

\def\calO{{\cal O}}
\def\calC{{\cal C}}
\def\calE{{\cal E}}
\def\calT{{\cal T}}
\def\calM{{\cal M}}
\def\calN{{\cal N}}
\def\calF{{\cal F}}
\def\calS{{\cal S}}
\def\calY{{\cal Y}}
\def\calV{{\cal V}}
\def\ibar{{\overline{\imath}}}
\def\chibar{{\overline{\chi}}}
\def\ttwo{{\vartheta_2}}
\def\tthree{{\vartheta_3}}
\def\tfour{{\vartheta_4}}
\def\ttwob{{\overline{\vartheta}_2}}
\def\tthreeb{{\overline{\vartheta}_3}}
\def\tfourb{{\overline{\vartheta}_4}}
\def\Str{{{\rm Str}\,}}

\def\bfell{{\boldsymbol \ell}}
\def\xx{\hspace{0.3cm}}
\def\xxl{\hspace{0.295cm}}
\def\xxh{\hspace{0.242cm}}
\def\yy{\hspace{0.115cm}}
\def\yyr{\hspace{-0.02cm}}

\def\qbar{{\overline{q}}}
\def\mm{{\tilde m}}
\def\nn{{\tilde n}}
\def\rep#1{{\bf {#1}}}
\def\ie{{\it i.e.}\/}
\def\eg{{\it e.g.}\/}

\newcommand{\newc}{\newcommand}
\newc{\gsim}{\lower.7ex\hbox{$\;\stackrel{\textstyle>}{\sim}\;$}}
\newc{\lsim}{\lower.7ex\hbox{$\;\stackrel{\textstyle<}{\sim}\;$}}

\newcommand{\red}[1]{\textcolor{red}{#1}}

\hyphenation{su-per-sym-met-ric non-su-per-sym-met-ric}
\hyphenation{space-time-super-sym-met-ric}
\hyphenation{mod-u-lar mod-u-lar--in-var-i-ant}


\def\inbar{\,\vrule height1.5ex width.4pt depth0pt}

\def\IC{\relax\hbox{$\inbar\kern-.3em{\rm C}$}}
\def\IQ{\relax\hbox{$\inbar\kern-.3em{\rm Q}$}}
\def\IR{\relax{\rm I\kern-.18em R}}
 \font\cmss=cmss10 \font\cmsss=cmss10 at 7pt
\def\IZ{\relax\ifmmode\mathchoice
 {\hbox{\cmss Z\kern-.4em Z}}{\hbox{\cmss Z\kern-.4em Z}}
 {\lower.9pt\hbox{\cmsss Z\kern-.4em Z}} {\lower1.2pt\hbox{\cmsss
 Z\kern-.4em Z}}\else{\cmss Z\kern-.4em Z}\fi}

\long\def\@caption#1[#2]#3{\par\addcontentsline{\csname
  ext@#1\endcsname}{#1}{\protect\numberline{\csname
  the#1\endcsname}{\ignorespaces #2}}\begingroup \small
  \@parboxrestore \@makecaption{\csname
  fnum@#1\endcsname}{\ignorespaces #3}\par \endgroup}
\catcode`@=12

\input epsf

\section{Introduction:  Background, motivation, and overview}

Recent data from the LHC seems to hint against the most minimal version of supersymmetry (SUSY), and there has been 
much recent interest in alternatives. It therefore seems a good moment to ask what the implications are for string phenomenology, given that 
SUSY is such an apparently integral part of string theory. 

Perhaps the major hurdle in constructing string models without spacetime SUSY
has been that they are generally {\it unstable}\/, with non-vanishing dilaton tadpoles.
However, in recent work~\cite{Abel:2015oxa}, we have been able to demonstrate the existence
of non-supersymmetric perturbative heterotic string models with exponentially suppressed dilaton tadpoles.
Such models are therefore virtually free of dilaton-related stability problems, and from 
this standpoint are similar to their SUSY cousins.
Moreover, models within this class can even be constructed which resemble 
the Standard Model or its various unified extensions. 
The existence of such models thus establishes a
starting point for the development of an entirely {\it non-supersymmetric}\/ string phenomenology ---
a phenomenology for possible physics beyond the Standard Model which is non-supersymmetric at any energy scale 
but which nevertheless descends directly from string theory.   As such, these models are UV complete,    
with finiteness ensured through entirely stringy mechanisms 
(such as modular invariance and ``misaligned SUSY''~\cite{missusy,supertraces,heretic}) which ensure finiteness even 
without SUSY.
This talk is dedicated to outlining the basic ideas behind these developments, with details to be found
in  Ref.~\cite{Abel:2015oxa}.

There are many reasons why the phenomenology of such non-supersymmetric models is of interest.
Clearly the major stumbling block for phenomenology at the electro-weak scale is the gauge hierarchy problem, namely the question of how to protect the electro-weak scale against quantum corrections from the UV completion of the theory. There are many symmetry-based ideas that have been explored within field theory towards solving this problem. However it is known that string theory provides additional symmetries --- modular invariance and misaligned SUSY, for example --- that cannot be seen within an effective field theory, except insofar as they might lead to approximate symmetries (such as non-compact shift symmetries). Moreover, almost all field-theoretic explorations of this subject (barring perhaps ones based on asymptotic safety) lack the very UV completion that one is trying to protect against. Therefore a successful explanation within effective field theory must be able to shield against {\it any} UV completion, regardless of its properties.  Supersymmetry is remarkable in that it protects against UV completions of any kind.  However it is important to ask if UV-complete but non-supersymmetric theories might provide additional (perhaps subtler) answers to the hierarchy problem. 

The results of Ref.~\cite{Abel:2015oxa} demonstrate that hierarchically separated scales {\it can} be natural within the context of non-supersymmetric string theories.  Note that we when refer to non-supersymmetric string theories, we do not mean theories that have spacetime SUSY at the Planck scale, with the SUSY broken in the low-energy field theory via purely field-theoretic mechanisms such as, for example, gaugino condensation.   Rather, we are referring to theories which are non-supersymmetric at all scales, including their fundamental Planck scales.  In such models, whatever supersymmetry might have otherwise existed has been 
destroyed through purely string-theoretic 
steps in the primordial model-construction process,
such as through 
particular choices of SUSY-breaking compactifications from ten dimensions 
that nevertheless respect modular invariance.
Such compactifications are often
generalised versions of Scherk-Schwarz compactification, and with a slight abuse of terminology one 
often speaks of breaking the SUSY ``spontaneously'' via such compactifications. 
The important point, however, is that only these purely string-theoretic construction methods 
ensure that the resulting string theory remains UV-finite with or without supersymmetry. 
It is modular invariance which ensures that the fundamental domain for the one-loop integral avoids those regions corresponding to the UV divergences of field theory. More physically, at the level of the string spectrum, this 
finiteness is ensured through a hidden so-called ``misaligned supersymmetry''~\cite{missusy,supertraces,heretic}
that always remains in the spectrum of any self-consistent string theory, even if supersymmetry itself is absent.
This misaligned SUSY corresponds to a subtle configuration of bosonic and fermionic states throughout the string spectrum in which no boson/fermion pairing exists, either exact or approximate, but which nevertheless conspires to produce
finite amplitudes.
Given this, the major accomplishment described in Ref.~\cite{Abel:2015oxa} 
has been to demonstrate that this UV-finiteness
can be realized within the context of string models which are phenomenologically semi-realistic 
and which can tolerate the large separations of scales that are needed for phenomenological purposes.

\begin{figure*}[h!]
\begin{center}
  \epsfxsize 5.5 truein \epsfbox {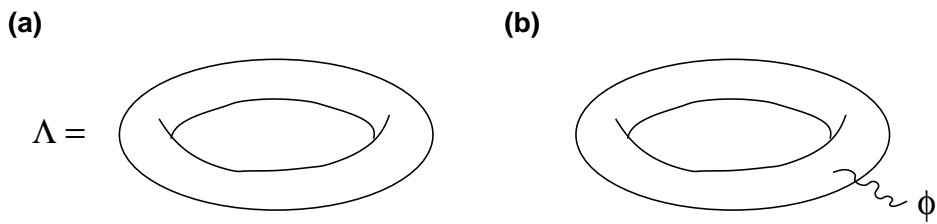}
\end{center}
\caption{(a)  The one-loop Casimir energy (cosmological constant) $\Lambda$.
         (b)  The one-loop one-point dilaton ``tadpole'' diagram.   In general, the value of the dilaton tadpole 
   is always proportional to $\Lambda$.  As a result, a non-zero cosmological constant
   implies a non-vanishing one-loop dilaton tadpole diagram, in turn indicating
  a linear term $\sim \phi$ in the effective potential.
This figure, like all figures in this talk, is adapted from Ref.~\cite{Abel:2015oxa}.}
\label{tadpole}
\end{figure*}

The first step in the hunt for such 
models is to tackle the issue of dilaton stability.  
It turns out that to one-loop order, the problem of dilaton stability is related to the problem of the 
cosmological constant, as sketched in Fig.~\ref{tadpole}.
This is encouraging, since 
in any non-supersymmetric string theory 
the cosmological constant is the first quantity that one would like to make hierarchically smaller than its generic value.
Thus, only those special theories in which the cosmological constant vanishes to leading order have a chance of being consistently stabilised.

The string models in 
constructed in Ref.~\cite{Abel:2015oxa} have this property as well.
Specifically, the models in 
Ref.~\cite{Abel:2015oxa} 
have exponentially suppressed one-loop cosmological constants.
Their dilatons are thus essentially stable, at least to one-loop order.
It is for this reason that these
models represent suitable platforms upon which to build a study
of non-supersymmetric string phenomenology.

The rest of this talk will be devoted to a discussion of these models
and their properties.
However, I should mention the many other works that have also considered 
non-supersymmetric string theories, beginning with the works
that originally adapted the Scherk-Schwarz mechanism to string theory.   
These include the original studies of the ten-dimensional $SO(16)\times SO(16)$ heterotic string~\cite{SOsixteen},
studies of the one-loop cosmological constants of non-supersymmetric strings~\cite{Rohm,nonSUSYgauge,Itoyama:1986ei,Itoyama:1987rc,Moore,
 Dienes:1990ij,KutasovSeiberg,missusy,supertraces,Kachru:1998hd,KachSilvothers,
  Shiu:1998he,Iengo:1999sm,DhokerPhong,Faraggi:2009xy},
their their finiteness properties~\cite{missusy,supertraces,Angelantonj:2010ic},
and their strong/weak coupling duality symmetries~\cite{Bergman:1997rf,julie1,julie2,Faraggi:2007tj}.
There have even been studies of the landscapes of such strings~\cite{Dienes:2006ut,Dienes:2012dc}. 
All studies of strings at finite temperature are also implicitly studies
of non-supersymmetric strings~(for early work in this area, 
see, \eg, Refs.~\cite{finitetemp,AtickWitten,wasKounnasRostand,Kounnas:1989dk,earlystringpapersfiniteT}).
In general, the non-supersymmetric string models 
which were studied were either non-supersymmetric by construction
or exhibited the same kind of spontaneous supersymmetry 
breaking that forms the basis of our work~\cite{Rohm, Ferrara:1987es, Ferrara:1987qp, Ferrara:1988jx, Kiritsis:1997ca, Dudas:2000ff,
Scrucca:2001ni, Borunda:2002ra, Angelantonj:2006ut}, achieved through
a stringy version of the Scherk-Schwarz mechanism~\cite{scherkschwarz} ---
indeed, potentially viable models within this class were constructed
in Refs.~\cite{Lust:1986kj, Lerche:1986ae,Lerche:1986cx, nonSUSYgauge, Chamseddine:1988ck, Font:2002pq,
Faraggi:2007tj,
Blaszczyk:2014qoa,
Angelantonj:2014dia}.
Non-supersymmetric string models have also been 
explored in a wide variety of other configurations~\cite{othernonsusy, Sagnotti:1995ga,Sagnotti:1996qj,Angelantonj:1998gj,Blumenhagen:1999ns,Sugimoto:1999tx,Aldazabal:1999tw,Angelantonj:1999xc,Forger:1999ev,Moriyama:2001ge,Angelantonj:2003hr,Angelantonj:2004yt,Dudas:2004vi,GatoRivera:2007yi}, including studies of the
relations between scales in various 
schemes~\cite{Antoniadis:1988jn, Antoniadis:1990ew, 
Antoniadis:1992fh,Antoniadis:1996hk,Benakli:1998pw,Bachas:1999es,Dudas:2000bn}. 


\section{The general framework}


\subsection{Scherk-Schwarz compactification:  General properties}

The string models of Ref.~\cite{Abel:2015oxa} are constructed
through supersymmetry-breaking Scherk-Schwarz compactifications from higher dimensions.
As such, they therefore share certain properties which are common to all models in which
spacetime supersymmetry is broken in this manner.

As mentioned, models in this class are  said to have ``spontaneously broken supersymmetry'': they have, for example, an identifiable order parameter for the breaking, namely the compactification scale, which we shall refer to generically as $1/R$ where $R$ is the compactification radius.  (In general, this scale will of course depend on all the moduli that describe the compactification manifold;  however this simplification will be sufficient for our purposes.)  Likewise, the cosmological constant  (which is essentially the vacuum energy density or Casimir energy density) {\it generically} goes like $1/R^4$.  Similarly, the Kaluza-Klein (KK) modes are typically split non-supersymmetrically 
with a splitting scale $1/2R$.  This, for example, is the mass of the gravitino in any of the theories we construct.

These properties notwithstanding,
it is important to emphasise from the 
outset that the spectrum of the resulting theory is truly non-supersymmetric at all energy scales. 
In particular, there is no scale beyond which the spectrum appears to be effectively supersymmetric. 
The winding modes of the theory have masses proportional to $R$ and so experience gross shifts in their masses 
which only increase with $R$.  The same is true at small radius, with winding modes and KK modes interchanged, and 
in the $R\rightarrow 0$ limit 
such theories typically remain non-supersymmetric but 
become effectively higher-dimensional, just as they do for $R\to \infty$.
Therefore, the distinction between supersymmetric and non-supersymmetric string theories
is {\it not}\/ merely a question of the {\it energy scale}\/ at which supersymmetry is broken, and 
as discussed more fully in Ref.~\cite{Abel:2015oxa} it would be wrong to view non-supersymmetric string models
as having been supersymmetric at high energy scales but subsequently subjected to some sort of SUSY-breaking mechanism
at lower energies. 
That the entirely supersymmetric theory reached at large $R$ is an extra-dimensional one is another indication of this fact: the gravitino and gaugino masses are the same order as the KK masses, so there is no scale at which 4D broken supersymmetry provides a good description of the phenomenology.

That said, it {\it does} make sense, at least partially, to speak of an 
effective spontaneously broken supersymmetric field theory, at the lowest orders of perturbation
theory. The heavy string modes provide a threshold contribution to the effective field theory which, thanks to the miracle 
of UV completion, is indeed finite and well behaved. 
Indeed, as discussed more fully in Ref.~\cite{Abel:2015oxa},
these properties persist --- even without supersymmetry --- as the result of a hidden ``misaligned supersymmetry'' which
neverthless remains in the spectrum 
of any tachyon-free, non-supersymmetric closed string model.

The nett result, then, is a theory in which the supersymmetry-breaking terms can be dialed to any value, even to the string scale itself, with non-supersymmetric threshold effects (such as, e.g., violations of the non-renormalisation theorem and hard supersymmetry-breaking operators) becoming more pronounced as the supersymmetry-breaking approaches the string scale. In this way the Scherk-Schwarz mechanism allows us to parametrically deform the theory away from one with a supersymmetric content towards an entirely non-supersymmetric one. This property of interpolation is an integral and 
important feature of our construction.

\subsection{The importance of interpolation: Proto-gravitons and their contributions to $\Lambda$}

In order to understand the importance of this interpolation property,
let us consider the various contributions to radiative 
corrections.
In general there can be many different kinds of physical and unphysical states which contribute to the one-loop partition function,
\beq
  Z(\tau) ~=~ {\tau_2}^{1-D/2} \sum_{m,n} \, a_{mn} \, \qbar^m q^n~,
\label{partfunctgen}
\eeq
where $q=e^{2\pi i \tau}$ in the usual nome, $\tau_1\equiv {\rm Re}\,\tau$, $\tau_2\equiv {\rm Im}\,\tau$, and $a_{nm}$ counts the number of bosons minus fermions with right- and left-moving worldsheet energies $(H_R,H_L) = (m,n)$.
Level-matched states with $m=n$ are physical,
while those with $m\not=n$ are ``unphysical'' and contribute only inside loops.
However, {\it every non-supersymmetric string model
necessarily contains off-shell tachyonic states with $(m,n)=(0,-1)$\/, leading to $a_{0,-1}\not=0$}.~
This is a theorem~\cite{Dienes:1990ij} 
which holds regardless of the specific class of non-supersymmetric string 
model under study and regardless of the particular GSO projections that might be imposed. 

It is easy to understand the origin of these states and their effect on the partition function.
We know that every string model contains 
a completely NS/NS sector from which the gravity multiplet arises:
\beq
        \hbox{graviton} ~~\subset ~~
              \tilde \psi_{-1/2}^\mu |0\rangle_R ~\otimes~ ~\alpha_{-1}^\nu |0\rangle_L~.
\label{graviton}
\eeq
Here $|0\rangle_{R,L}$ are the right- and left-moving vacua of the heterotic string, 
$\tilde \psi_{-1/2}^\mu$ represents the excitation
of the right-moving world-sheet Neveu-Schwarz fermion $\tilde \psi^\mu$,
and $\alpha_{-1}^\nu$ represents the excitation
of the left-moving coordinate boson $X^\nu$.
Indeed, no self-consistent GSO projection can possibly 
eliminate this gravity multiplet from the string spectrum.
However, given that the graviton is always in the string spectrum, 
then there must also exist in the string spectrum a corresponding state
for which the left-moving coordinate oscillator is {\it not}\/ excited:
\beq
        \hbox{proto-graviton:}~~~~~~~~~~~~
              \tilde \psi_{-1/2}^\mu |0\rangle_R ~\otimes~ |0\rangle_L~.
\label{protograviton}
\eeq
This ``proto-graviton'' state has world-sheet energies $(E_R,E_L)=(m,n)=(0,-1)$, and is thus off-shell  
and tachyonic.  Nevertheless, it is always there in the string spectrum along with the graviton.

Normally one ignores such things in phenomenology, firstly because they cannot appear as asymptotic states in 
any scattering (hence they are referred to as  ``unphysical'' which we consider to be something of a misnomer), and secondly because, in a supersymmetric theory, any contribution to the partition function from the proto-graviton is automatically cancelled  by an equal and opposite one from its superpartner, the proto-gravitino. In the context of non-supersymmetric strings, however, the latter is absent (or lifted to the 
SUSY-breaking scale). Thus we can quite generally write the first term in the $q$-expansion of {\it any} non-SUSY string theory. 
As evident from Eq.~(\ref{protograviton}),
the proto-graviton states transform as vectors under the transverse spacetime Lorentz symmetry $SO(D-2)$.
Thus, any non-supersymmetric string theory in $D$ uncompactified spacetime dimensions must have a partition function
which begins with the contribution 
\beq
          Z(\tau) ~=~ {D-2 \over q } ~+~ ...
\label{protogravitoncontributions}
\eeq
One may also easily evaluate the contributions that the various states make to the cosmological constant. 
These are given by
\beq
         \Lambda^{(D)} ~\equiv~
     -\half \,{\cal M}^D\, \int_{\cal F} {d^2 \tau\over {\tau_2}^2}
             Z(\tau)~
\label{cosconstdef}
\eeq
where ${\cal M}$ is the reduced string scale
and 
\beq
   {\cal F}~\equiv ~\lbrace \tau:  ~|{\rm Re}\,\tau|\leq \half,~
 {\rm Im}\,\tau>0, ~|\tau|\geq 1\rbrace
\label{Fdef}
\eeq
is the fundamental domain
of the modular group.

As a toy example, and also to illustrate the general structure of interpolating models, consider 
 a $D=10$ theory compactified on a twisted circle. Any such $(D-1)$-dimensional model has a partition function that takes the general form~\cite{Rohm,finitetemp,AtickWitten,wasKounnasRostand,Kounnas:1989dk}
\beqn
        Z_{\rm string}(\tau,R)  &=&
           Z^{(1)}(\tau) ~ \calE_0(\tau,R) ~+~
           Z^{(2)}(\tau) ~ \calE_{1/2}(\tau,R) \nonumber\\
           && ~~+~  Z^{(3)}(\tau) ~ \calO_{0}(\tau,R) ~+~
           Z^{(4)}(\tau) ~ \calO_{1/2}(\tau,R) ~
\label{EOmix}
\eeqn
where the $\calE_{0,1/2}$ and $\calO_{0,1/2}$ functions indicate various restricted summations over KK and winding modes, as described in Sect.~III.B of Ref.~\cite{Abel:2015oxa}.
The models are considered to be ``interpolating'' between two different ten-dimensional models $M_1$ and $M_2$ as
$R\to\infty$ and $R\to 0$, respectively, with
$Z^{(1)} + Z^{(2)}$ reproducing the  partition function of $M_1$ and
$Z^{(1)} + Z^{(3)}$ reproducing the partition function $M_2$.
In this talk we shall be concerned with situations in which $M_1$ has spacetime supersymmetry  
but $M_2$ does not.
Thus $Z^{(2)}= -Z^{(1)}$, so that there are only three different sectors making non-supersymmetric
contributions to the cosmological constant:
$\calE_0-\calE_{1/2}$, $\calO_0$, and $\calO_{1/2}$.

\begin{figure*}[t!]
\begin{center}
  \epsfxsize 2. truein \epsfbox {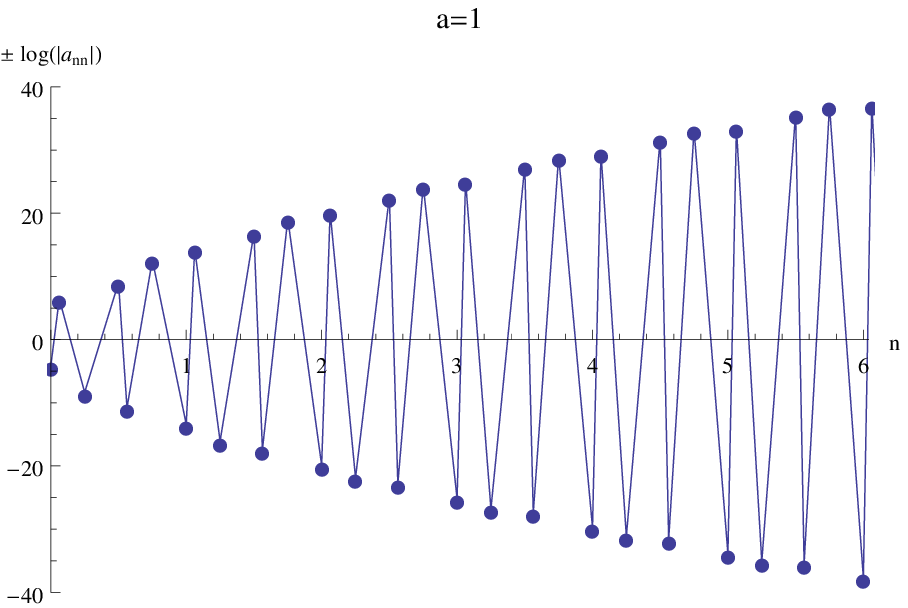}    
\hskip 0.2 truein
  \epsfxsize 2. truein \epsfbox {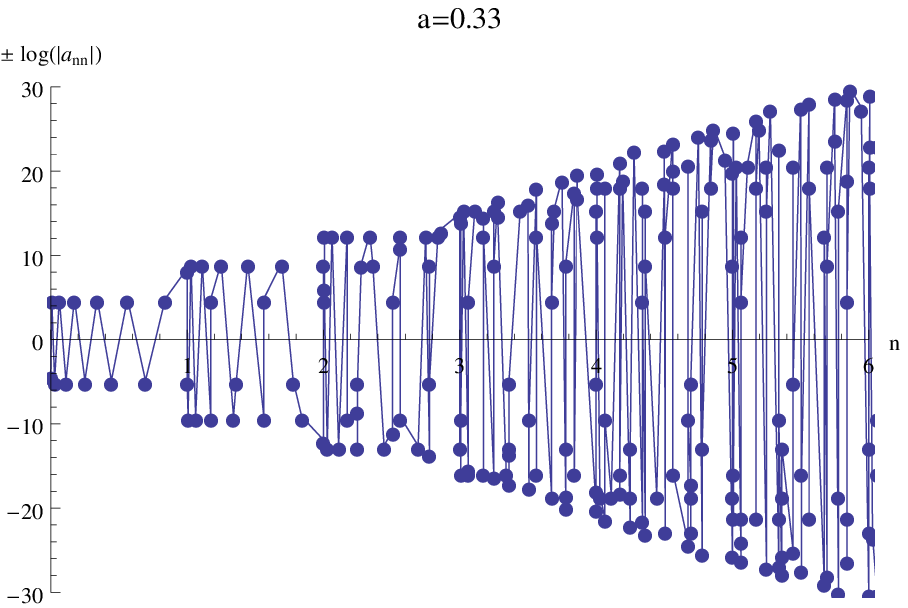}
\end{center}
\begin{center}
  \epsfxsize 2. truein \epsfbox {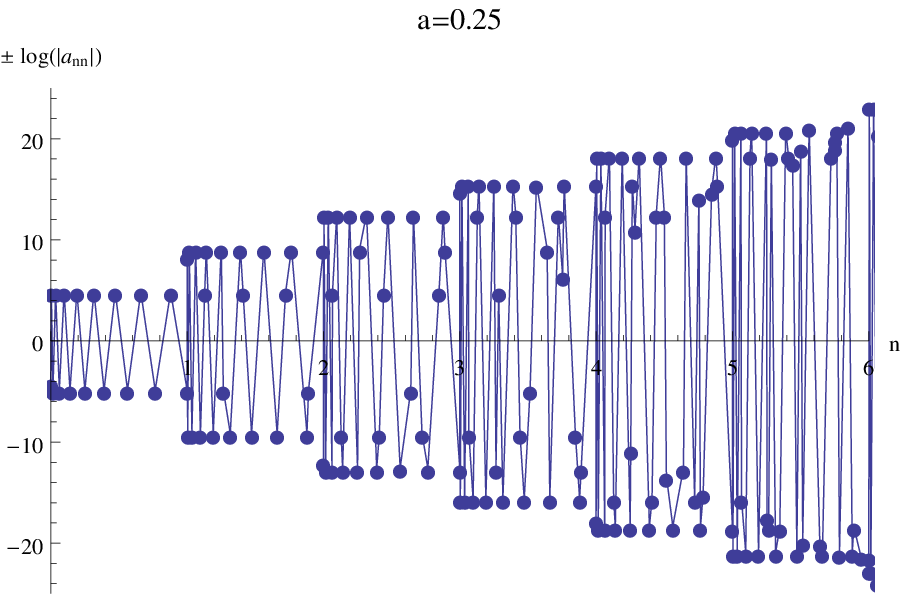}
\hskip 0.2 truein
  \epsfxsize 2. truein \epsfbox {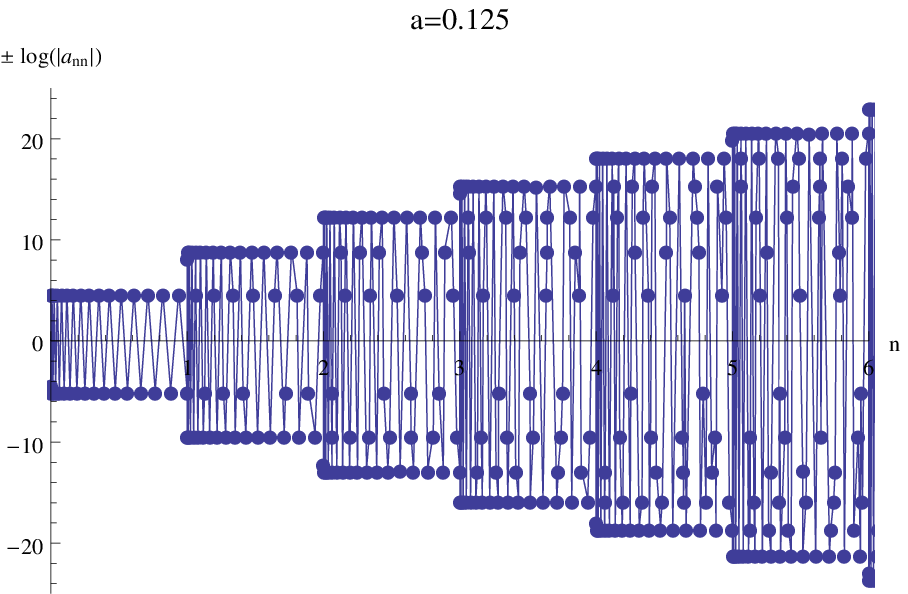}
\end{center}
\caption{Degeneracies of physical states for a particular 9D interpolating model
described in Ref.~\cite{Abel:2015oxa}.
Results are shown for  
$a=1$ (upper left), $a=0.33$ (upper right), $a=0.25$ (lower left), $a=0.125$ (lower right),
where $a\equiv \sqrt{\alpha'}/R$.    
Within each plot, points are connected in order of increasing world-sheet energy $n$.  In all
cases we see that surpluses of bosonic states alternate with surpluses of fermionic states
as we proceed upwards in $n$;  this behavior is the signal of an underlying ``misaligned  
supersymmetry''~\cite{missusy,supertraces,heretic} 
which exists within all modular-invariant non-supersymmetric tachyon-free 
string theories and which is ultimately responsible for the finiteness of closed strings  --- even 
in the absence of spacetime supersymmetry.  For $R=\sqrt{\alpha'}$ (or $a=1$), we see that this
oscillation between bosonic and fermionic surpluses occurs within 
the exponentially growing envelope function $|a_{nn}|\sim e^{c\sqrt{n}}$ associated
with a Hagedorn transition.
However, as the compactification radius increases (or
equivalently as $a\to 0$), we see that a hierarchy begins to emerge between the oscillator 
states and their KK excitations;
the oscillator states continue to experience densities of states which are exponentially growing as 
functions of $n$,  but 
their corresponding KK excitations are densely packed within 
each interval $(n,n+1)$ and, as expected, exhibit constant state degeneracies.}
\label{fig:interplots}
\end{figure*}

The bose-fermi non-degeneracies (i.e., the values of $a_{nn}$ as a function of $n$) 
are shown in Fig.~\ref{fig:interplots} for a specific model
in this class, for different values of the dimensionless inverse radius $a\equiv \sqrt{\alpha'}/R$.    
This spectral information clearly illustrates the 
non-softness of the supersymmetry breaking in such models, 
namely the fact that no matter how much one might attempt to ``restore'' the supersymmetry by 
increasing the radius $\sqrt{\alpha'}/R=a\to 0$, 
the spectrum remains non-supersymmetric with $a_{mn}\not =0$ for all $(m,n)$
as long as $R$ remains finite.
Despite this feature, for large but finite $R$ the low-lying spectrum resembles that of a broken-SUSY higher-dimensional theory
(with towers of KK modes and their slightly displaced would-be superpartners), 
while the intermediate and heavy spectra are more violently non-supersymmetric and thereby produce
important non-supersymmetric threshold effects. 

Within such interpolating models, the contributions to the cosmological constant in the
$\sqrt{\alpha'}/R=a\to 0$ limit from a given state with world-sheet energies $(m,n)$
in the different $\calE/\calO$-sectors are found~\cite{Abel:2015oxa} to be as follows: 
\beq
\begin{tabular}{||c|c|| c ||}
\hline
\hline
~~ {\rm sector}~~ & ~~{\rm state}~~& ~~contribution~to~$\Lambda$~~\\
\hline
\hline
~~$\calE_0-\calE_{1/2}$~~ & ~~ $m=n=0$~~ & ~~$-\displaystyle{{[4 (D/2-1)!/ \pi^{D/2}}] \, a^{D-1} }$ \\[5 pt]
\hline
~~$\calE_0-\calE_{1/2}$~~ & ~~ $m=n\not= 0$~~ & ~~$\displaystyle{4 (2\sqrt{m} a)^{(D-1)/2} e^{-4\pi \sqrt{m}/a}}$~~\\
\hline
~~$\calE_0-\calE_{1/2}$~~ & ~~ $m\not =n$~~ & ~~$\displaystyle{-[{4\sqrt{2} / \pi]} e^{-2\pi (m+n)} a^2 e^{-\pi/a^2}}$~~\\
\hline
~~$\calO_{0,1/2}$~~ & ~~ any~$(m,n)$~~ & ~~ $\displaystyle{{[2\sqrt{2} / \pi]} e^{-2\pi (m+n)} a^2 e^{-\pi/a^2}}$~~\\
\hline
\hline
\end{tabular}
\label{tableofcontributions}
\eeq
In this table, $D$ represents the dimensionality of the theory in question {\it prior}\/ 
to the compactification on the twisted circle. 
At large radii, the leading contribution to the cosmological constant is given by the nett contribution coming from the massless $m=n=0$ physical states.  As evident from Table~\ref{tableofcontributions}, this contributions takes the form
\beq
  \Lambda ~\sim~ (N_b^{(0)}-N_f^{(0)}) \, a^{D-1}~+~...
\eeq
where $N_b^{(0)}$ and $N_f^{(0)}$ are respectively the numbers of bosonic/fermionic
states which remain massless in our theory after SUSY-breaking has already occurred.
In more general compactifications from $D$ down to $d$ dimensions one 
would find 
\begin{equation} 
       \Lambda_d ~\sim~ (N_b^{(0)}-N_f^{(0)}) \, a^{d}~+~ ... ~,
\end{equation}
thereby recovering the same contribution to the Casimir energy that one would infer in 
a higher-dimensional field theory. 
Quite remarkably, however,
we see from 
Table~\ref{tableofcontributions} that all other contributions to the cosmological constant
are exponentially suppressed! 
Thus, if we can construct models for which $N_b^{(0)}=N_f^{(0)}$,
we will have succeeded in constructing string models for which the cosmological 
constant is {\it exponentially}\/ suppressed.

\begin{figure*}[t!]
\begin{center}
  \epsfxsize 6.0 truein \epsfbox {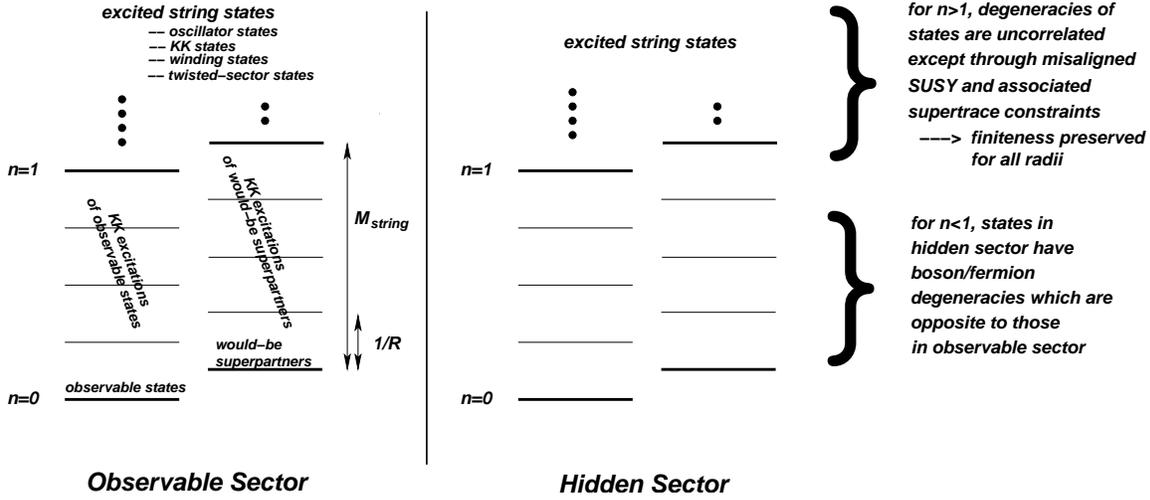}
\end{center}
\caption{The structure of the spectrum of a generic interpolating model with suppressed
cosmological constant in the limit of large interpolating radius.
States with masses below $M_{\rm string}$ (or below $n=1$)  consist of massless observable states,
massless hidden-sector states,
their would-be superpartners, and their lightest KK excitations.
For these lightest states,
the nett (bosonic minus fermionic) numbers of degrees of freedom 
from the hidden sector are exactly equal and opposite
to those from the observable sector for all large radii. 
Note that this cancellation of nett physical-state degeneracies between the observable and hidden sectors
bears no connection with any supersymmetry, either exact or approximate, in the string spectrum.
Nevertheless, it is this conspiracy between the observable and hidden sectors which suppresses
the overall cosmological constant and enhances the stability of these strings.
For the heavier states, by contrast, the observable and hidden sectors need no longer 
supply equal and opposite numbers of degrees of freedom. Nevertheless the entire theory remains finite at one-loop order
through the misaligned supersymmetry~\cite{missusy,supertraces,heretic} illustrated 
in Fig.~\protect\ref{fig:interplots}.}
\label{fig:cdcspectra}
\end{figure*}

We emphasize that having $N_b^{(0)}=N_f^{(0)}$ is only a statement about massless states.
Moreover, this requirement does not require an actual supersymmetry, even at the massless level;   all that is required 
are equal numbers of massless bosonic and fermionic degrees of freedom.
Finally, not all of these degrees of freedom are required to exist in a visible sector --- 
some may carry quantum numbers that correspond to visible-sector states,
while others may carry quantum numbers that place them in a hidden sector.
Neither sector by itself needs to  
exhibit $N_b^{(0)}=N_f^{(0)}$ as long as both sectors together combine to maintain this relation.
The generic structure of the low-lying spectrum for such models is illustrated
in Fig.~\ref{fig:cdcspectra}.

It is important to note that when 
$N_b^{(0)}=N_f^{(0)}$,
the one-loop cosmological constant is exponentially suppressed  {\it even when the contributions from the unphysical states 
are included}.
Note that the contribution from such states is independent of the number of spacetime dimensions $D$. 
One can understand this from the fact that $D$-dependence 
requires states to be able to propagate long distances and thereby ``feel'' the full spacetime --- something
 which unphysical states are not able to do.
Nevertheless these states still contribute to the cosmological constant because the bottom of the fundamental domain (i.e., the UV end of the one-loop integral) is curved.
Indeed, as noted in Ref.~\cite{Abel:2015oxa}, the contribution from the proto-graviton states with $(m,n)=(0,-1)$ exceeds that of even the massless $(0,0)$ physical 
contributions for $a\lsim 0.54$. Thus such interpolating models with $R\lsim 2$ have little chance of 
being stable.  

All of these observations
render the cleanest assumptions about the compactification scale (namely that $M_c\equiv 1/R \approx  M_{\rm string}$) problematic.
Indeed, in such models it is not always clear how to separate oscillator states from KK states
and/or winding states;  there even exist examples of such models which transcend the 
notion of having a compactification geometry altogether and in which no 
compactification geometry can be identified.
For such models we typically obtain a cosmological constant of order $\Lambda\sim M_{\rm string}$.
Of course, even within such string models, there remains the possibility that $\Lambda$ might still
vanish through some other mechanism.  For example, the proposals in 
Refs.~\cite{Moore,Dienes:1990ij,Kachru:1998hd} all rely on different kinds of symmetry arguments 
for cancelling $\Lambda$ within closed string models for which $M_c\sim M_{\rm string}$.
Unfortunately, no string models have ever been constructed exhibiting the symmetries 
proposed in Refs.~\cite{Moore,Dienes:1990ij}, and the 
mechanism proposed in Ref.~\cite{Kachru:1998hd} may actually fail at higher loops~\cite{Shiu:1998he,Iengo:1999sm}.

The alternative possibility that we are proposing is to consider models in which $M_{\rm string}$ is fixed
but $M_c$ is taken to be a free, adjustable variable.
Indeed, we can go even further and imagine that our compactification volume is characterized
by many different compactification scales $M_c^{(i)}$, each of which we might consider a free parameter;
such a scenario would emerge, for example, if our $d$-dimensional compactification manifold is a $d$-torus
with different radii of compactification $R_i$, $i=1,...,d$.
In general, as the volume of compactification $V_d$ is taken to infinity, we effectively produce a
string model in $d$ additional spacetime dimensions.  This higher-dimensional model is the general equivalent of 
our $M_1$ above.  For closed strings, T-duality then ensures that we also 
produce a model in $d$ additional spacetime dimensions
as $V_d\to 0$, which is the equivalent of $M_2$.  The model with variable compactification volumes
can thus be said to {\it interpolate}\/ between the two higher-dimensional endpoint models, $M_1$ and $M_2$, in the 
same manner as the interpolation between the two ten-dimensional models above. 

Such interpolating models offer a number of distinct advantages when it comes to suppressing the cosmological
constant.  Since the model $M_1$ is supersymmetric in our construction, we are assured that $\Lambda=0$ when $V_d\to \infty$.
Moreover, since $M_2$ is non-supersymmetric, the spacetime supersymmetry is 
broken for all finite $V_d$.
It is therefore reasonable to assume that we can dial $V_d$ to a sufficiently large value in order to obtain
a cosmological constant of whatever size we wish.
Even more compellingly, however, there is a widespread belief that spacetime supersymmetry, if it exists at all in
nature, is broken at the TeV-scale, with superpartners having masses $\sim {\cal O}({\rm TeV}\/)$.
Indeed, as first suggested in Refs.~\cite{Rohm, Antoniadis:1990ew}, these sorts of scenarios 
with large compactification volumes
are relatively easy to incorporate with the interpolating-model framework
with $M_c\sim {\cal O}({\rm TeV})$.
Of course, there is absolutely no reason within our construction why this value for the radius should be chosen.

Perhaps even more importantly, we also note that within this construction, the {\it scale}\/ of the cosmological constant
is no longer tied to the effective scale of the supersymmetry breaking.
In particular, although we can consider the scale of supersymmetry
breaking in these models to be given by $M_c=1/R$, for models with 
$N_b^{(0)}=N_f^{(0)}$ 
the cosmological constant is exponentially suppressed, scaling as
${\cal O}(e^{-4\pi M_{\rm string}/M_c})$.
This allows $M_c$ to be even larger than ${\cal O}$(TeV)
without destroying the all-important stablity of these theories.
This additional stability is a  
precursor of what one might eventually hope to achieve for scalar masses.


\section{Semi-realistic interpolating string models with exponentially small cosmological constant}

\subsection{Outline of construction technique }

We now outline the construction of our stable semi-realistic non-supersymmetric string models.
For technical reasons it is advantageous to interpolate between $M_1$ and $M_2$ models in {\it six}\/ dimensions rather
than five. We therefore begin with six-dimensional models $M_1$ that have
 $\calN=1$ supersymmetry. Such models are most conveniently obtained by lifting to six dimensions 
semi-realistic four-dimensional 
$\calN=1$ string models, for example those in Refs.~\cite{Antoniadis:1989zy, Antoniadis:1990hb, 
  Faraggi:1991be, Faraggi:1991jr, Faraggi:1992fa, Faraggi:1994eu, Dienes:1995bx}
which are already on the market. Our objective is to retain as far as possible their desirable phenomenological features.

Once we have constructed such suitable models $M_1$, the next step is to compactify back down to four dimensions.
The four-dimensional $\calN=1$ model that results from
compactifying back to four dimensions on a $T_2/\IZ_2$ orbifold 
can be compared with the four-dimensional $\calN=0$ model that results from a coordinate-dependent compactification (CDC) 
on the same orbifold using the techniques of 
Refs.~\cite{Ferrara:1987es, Ferrara:1987qp, Ferrara:1988jx}. 

\begin{figure*}[t!]
\begin{center}
  \epsfxsize 6.0 truein \epsfbox{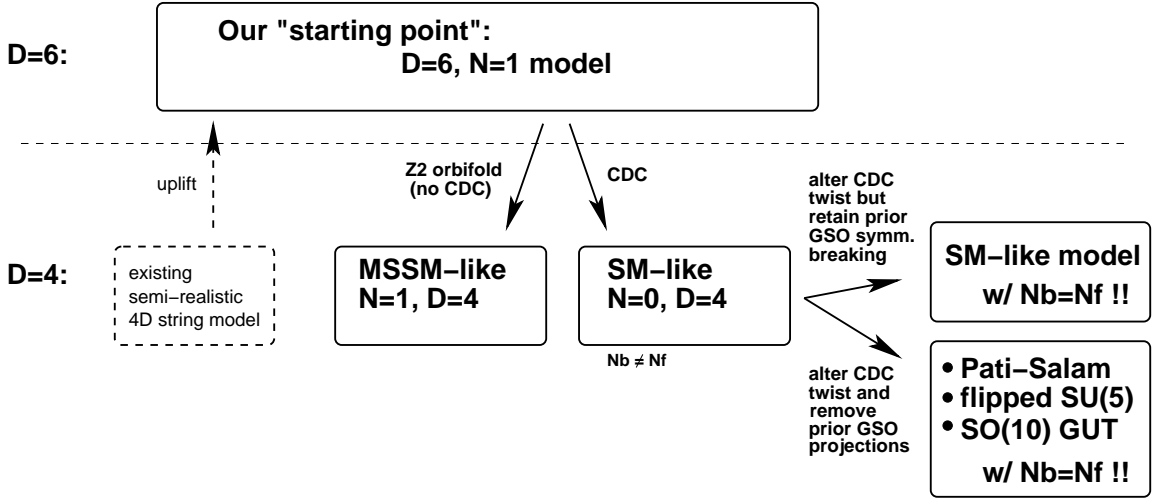}
\end{center}
\vspace{-1cm}
\caption{Roadmap illustrating our procedure for constructing 
semi-realistic non-supersymmetric string models with $N_b^{(0)}=N_f^{(0)}$, as discussed in the text.}
\label{fig:map}
\end{figure*}

Our final step is to take the resulting $\calN=0$ model and introduce modifications
to obtain $N_b^{(0)}=N_f^{(0)}$, as required to produce exponentially suppressed cosmological constant.
There are several different ways in which this can be done.
One way is to alter the final Scherk-Schwarz (CDC) twist but retain the prior GSO symmetry breaking:  this can produce 
an SM-like model. By contrast, altering the final twist and also removing prior GSO projections
can lead to a variety of additional models:  a  Pati-Salam-like model, a flipped-$SU(5)$ ``unified'' model,
and an $SO(10)$ ``unified'' model,  each also with $N_b^{(0)}=N_f^{(0)}$.
The procedure is outlined in Fig.~\ref{fig:map}.
Undoubtedly these models are only several within an entire new terrain 
which deserves exploration. 

\subsection{Example:  A stable, non-supersymmetric 4D Pati-Salam model}
\label{model2} 
As an illustration we present the Pati-Salam model mentioned above. 
 It is defined by the generalised GSO vectors 
\begin{eqnarray}
V_0&=& - {\scriptstyle\frac{1}{2}}[~11~111~111~ | ~1111~11111~111~11111111~]\nonumber\\ 
V_1&=& - {\scriptstyle\frac{1}{2}}[~00~011~011~ | ~1111~11111~111~11111111~]\nonumber\\ 
V_2&=& - {\scriptstyle\frac{1}{2}}[~00~101~101~ | ~0101~00000~011~11111111~]\nonumber\\ 
b_3&=& - {\scriptstyle\frac{1}{2}}[~10~{1}0{0}~{0}0{1}~ | ~0001~11111~001~10000111~]\nonumber\\ 
V_4&=& - {\scriptstyle\frac{1}{2}}[~00~101~101~ | ~0101~00000~011~00000000~]\nonumber\\ 
V_5&=& - {\scriptstyle\frac{1}{2}}[~00~0{0}{0}~0{1}{1}~ | ~0100~11100~000~11100111~]\nonumber\\ 
{\bf e} &=& ~~~{\scriptstyle\frac{1}{2}}[~00~101~101~ | ~1011~00000~000~00011111~]\, .
\end{eqnarray} 
where the notation is standard in fermionic string constructions  and is summarized in the Appendix of Ref.~\cite{Abel:2015oxa}.
The vector ${\bf e}$ shows the action of the CDC on the right-moving space-time world-sheet degrees of freedom 
(listed on the left above) and the 
left-moving internal degrees of freedom (listed on the right above). 
The vector dot products and $k_{ij}$ structure constants for this model are given by
\begin{equation}
V_i\cdot V_j =\left( \begin{array}{c}1\hspace{0.1cm}0\hspace{0.1cm}0\hspace{0.1cm}0\hspace{0.1cm}0\hspace{0.1cm}0\nonumber \\0\hspace{0.1cm}0\hspace{0.1cm}{\scriptstyle\frac{1}{2}}\hspace{0.1cm}{\scriptstyle\frac{1}{2}}\hspace{0.1cm}{\scriptstyle\frac{1}{2}}\hspace{0.1cm}0\nonumber \\0\hspace{0.1cm}{\scriptstyle\frac{1}{2}}\hspace{0.1cm}0\hspace{0.1cm}1\hspace{0.1cm}0\hspace{0.1cm}{\scriptstyle\frac{3}{2}}\nonumber \\0\hspace{0.1cm}{\scriptstyle\frac{1}{2}}\hspace{0.1cm}1\hspace{0.1cm}0\hspace{0.1cm}0\hspace{0.1cm}{\scriptstyle\frac{3}{2}}\nonumber \\0\hspace{0.1cm}{\scriptstyle\frac{1}{2}}\hspace{0.1cm}0\hspace{0.1cm}0\hspace{0.1cm}0\hspace{0.1cm}0\nonumber \\0\hspace{0.1cm}0\hspace{0.1cm}{\scriptstyle\frac{3}{2}}\hspace{0.1cm}{\scriptstyle\frac{3}{2}}\hspace{0.1cm}0\hspace{0.1cm}0\end{array}\right)\:\: 
             \mbox{ mod\, (2)}~,~~~~~~ \:\:\:
k_{ij} =\left( \begin{array}{c}0\hspace{0.1cm}0\hspace{0.1cm}0\hspace{0.1cm}{\scriptstyle\frac{1}{2}}\hspace{0.1cm}0\hspace{0.1cm}0\nonumber \\0\hspace{0.1cm}0\hspace{0.1cm}0\hspace{0.1cm}{\scriptstyle\frac{1}{2}}\hspace{0.1cm}0\hspace{0.1cm}0\nonumber \\0\hspace{0.1cm}{\scriptstyle\frac{1}{2}}\hspace{0.1cm}0\hspace{0.1cm}0\hspace{0.1cm}0\hspace{0.1cm}{\scriptstyle\frac{1}{2}}\nonumber \\{\scriptstyle\frac{1}{2}}\hspace{0.1cm}0\hspace{0.1cm}0\hspace{0.1cm}0\hspace{0.1cm}0\hspace{0.1cm}{\scriptstyle\frac{1}{2}}\nonumber \\0\hspace{0.1cm}{\scriptstyle\frac{1}{2}}\hspace{0.1cm}0\hspace{0.1cm}0\hspace{0.1cm}0\hspace{0.1cm}0\nonumber \\0\hspace{0.1cm}0\hspace{0.1cm}0\hspace{0.1cm}0\hspace{0.1cm}0\hspace{0.1cm}0\end{array}\right)\, .
\end{equation}
 The gauge-group structure is
\begin{equation} 
 G ~=~ SO(4)\otimes U(1)\otimes U(1)\otimes 
                 \underbrace{SO(6)\otimes SO(4)}_{\rm contains~SM}
               \otimes U(1)\otimes U(1)\otimes U(1)\otimes U(1)\otimes SO(4)\otimes SO(4)\otimes SO(6)\, , 
\label{PSgg}
\end{equation}
where the Pati-Salam group corresponding to the visible sector is indicated.
This model, which has four quasi-supersymmetric chiral generations of massless 
        untwisted matter but no twisted matter,
 has $N^{(0)}_b=N^{(0)}_f=416$ complex massless degrees of freedom in the untwisted sector. 
 The spectrum is shown in Tables~\ref{table:I} and \ref{table:J}.
Many similar examples can be found. 

\section{Phenomenological properties}

\subsection{Spectrum}

\begin{figure*}[t!]
\begin{center}
  \epsfxsize 2 truein \epsfbox {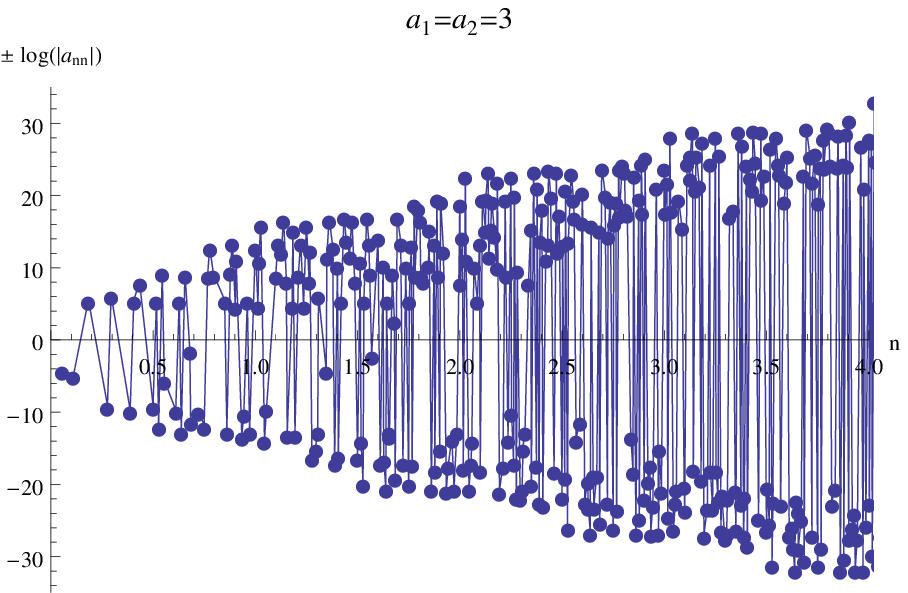}    
\hskip 0.2 truein
  \epsfxsize 2 truein \epsfbox {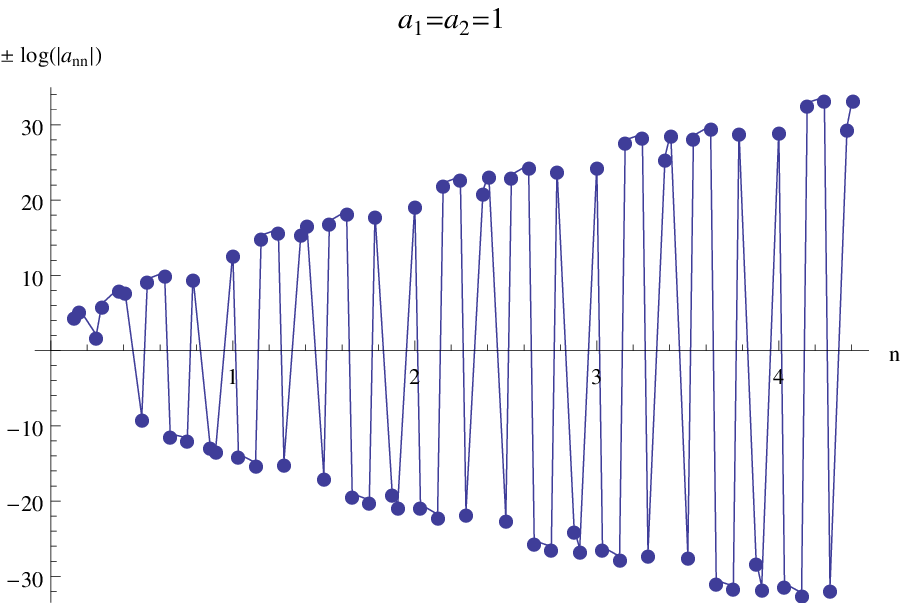}
\end{center}
\begin{center}
  \epsfxsize 2 truein \epsfbox {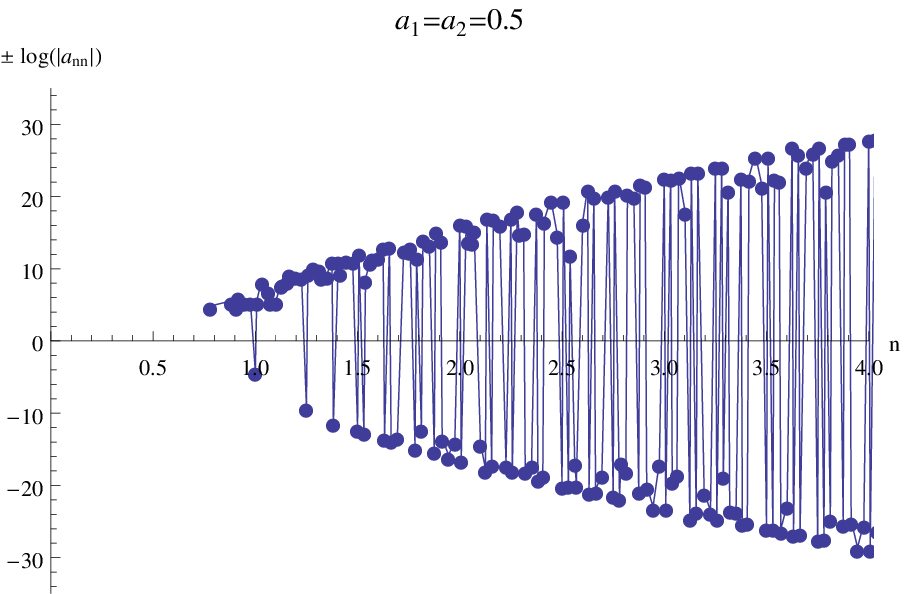}
\hskip 0.2 truein
  \epsfxsize 2 truein \epsfbox {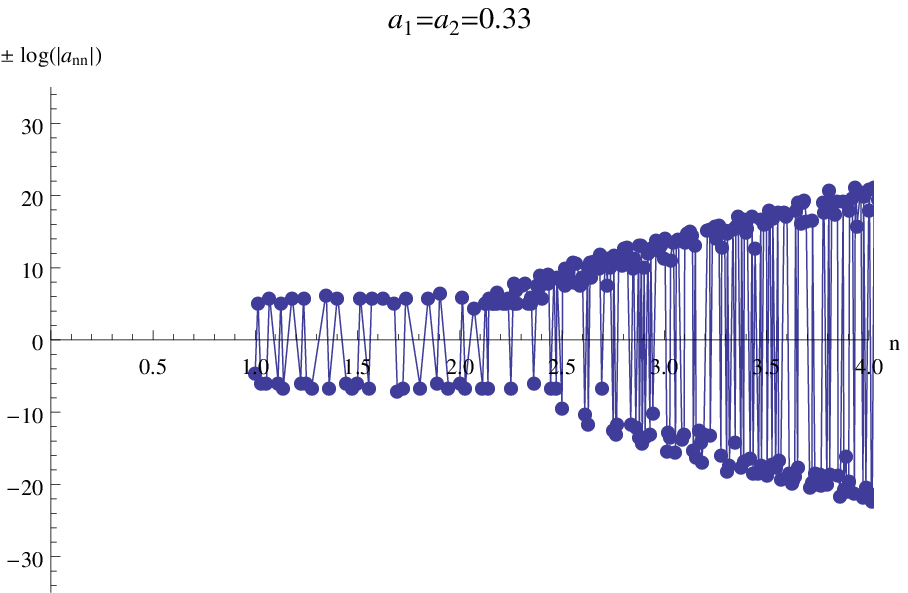}
\end{center}
\begin{center}
  \epsfxsize 2 truein \epsfbox {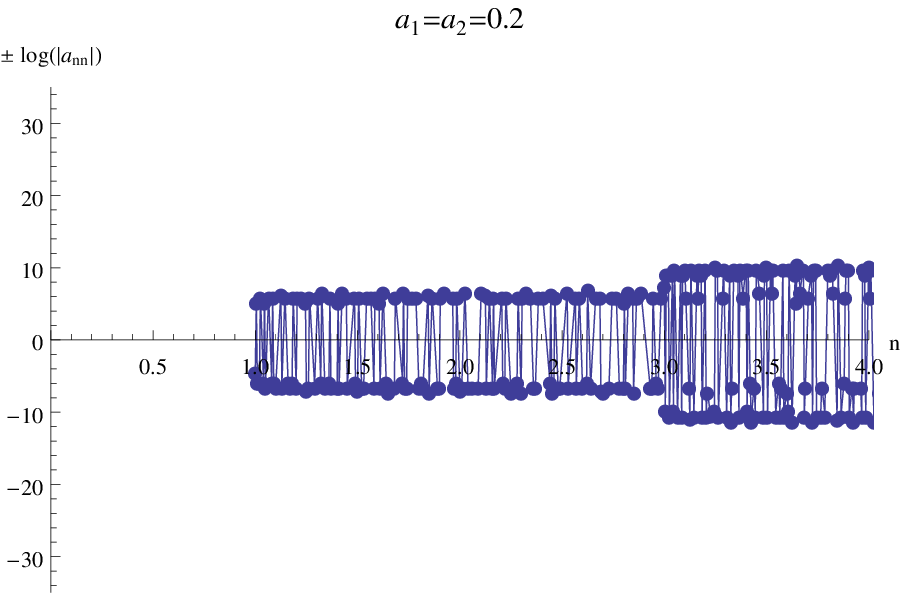}
\hskip 0.2 truein
  \epsfxsize 2 truein \epsfbox {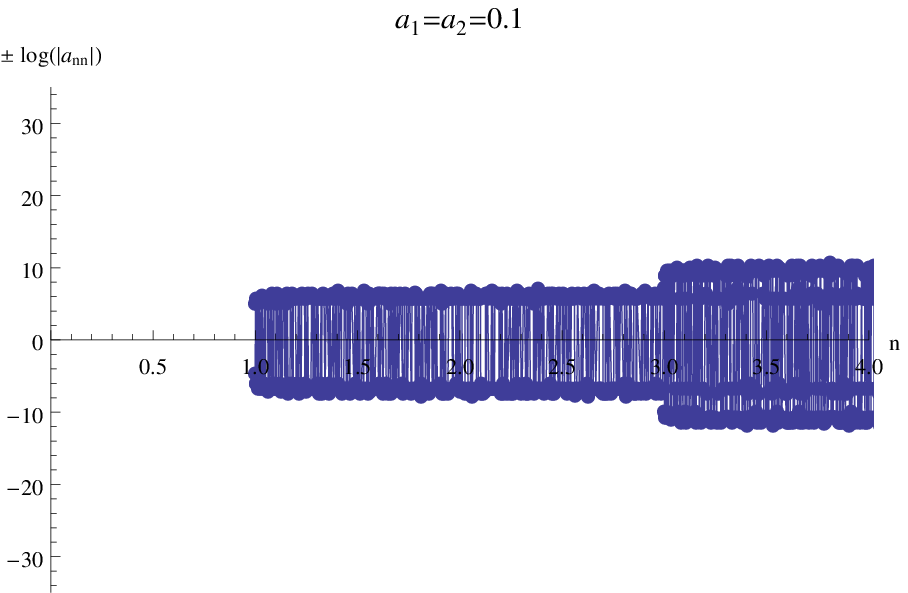}
\end{center}
\caption{Degeneracies of physical states for the Pati-Salam model 
with exponentially suppressed cosmological constant.
The inverse radius $a=\sqrt{\alpha'}/R$ varies from $a=3$ (upper left)
to $a=0.1$ (lower right).
Comparing with Fig.~\protect\ref{fig:interplots}, we see that all of the general features associated with  interpolating models  
survive, including 
a smoothly growing exponential envelope function for $a\sim {\cal O}(1)$ which slowly deforms into a discretely step-wise growing exponential function as $a\to 0$.  
This reflects the emerging hierarchy between KK states and oscillator states.
However, we also observe a critical new feature which reflects the fact that this model has an exponentially    
suppressed cosmological constant:  the removal or ``evacuation'' of all 
non-zero nett state degeneracies $a_{nn}$ for $n\leq 1$ for
sufficiently small $a$.  Thus, for sufficiently large radius, the spectrum of such models develops an exact 
boson/fermion degeneracy  
for all relevant mass levels $n<1$, even though there is no supersymmetry anywhere in the spectrum.
Indeed, as illustrated in Fig.~\protect\ref{fig:cdcspectra}, this degeneracy does {\it not}\/ occur through a pairing of states 
with their would-be superpartners, but rather as the result of the balancing of non-zero nett degeneracies
associated with a non-supersymmetric  {\it observable}\/ sector 
against the degeneracies associated with a non-supersymmetric {\it hidden}\/ sector.}
\label{fig:cdcmissusyplots}
\end{figure*}

The phenomenological structure of models such as those discussed above is very general. First, for large compactification
radii, the spectrum itself is arranged according to the characteristic form sketched in Fig.~\ref{fig:cdcspectra},
with visible and hidden sectors together conspiring to produce $N_b^{(0)}=N_f^{(0)}$.
This boson/fermion degeneracy holds only for the low-lying states, however, and disappears at higher mass levels.
This structure is also apparent in the plots of nett bose-fermi number in Fig.~\ref{fig:cdcmissusyplots}, which illustrate
the ``misaligned-supersymmetry'' properties of the above Pati-Salam model for a variety of different compactification radii.
Indeed, as the compactification radius increases beyond a certain critical value, we see from 
Fig.~\ref{fig:cdcmissusyplots}
that the entire KK spectrum begins to exhibit a bose-fermi degeneracy below the string scale --- all this despite the fact that the 
theory is completely non-supersymmetric at all mass levels and energy scales.

\begin{table}[H]
\centering
\begin{tabular}{|c|l|c|l|l|}
\hline
~Sector~ & ~~States remaining after CDC~~ & ~Spin~ & ~$SU(4)\otimes SU(2)_L\otimes SU(2)_R$~ & ~Particle~ \\
\hline
& & & &\\
\multirow{8}{*}{${V_0+V_2}$}  & $~|\alpha\rangle_R \otimes {\overline{\psi}}^i_0{\overline{\psi}}^a_0|\hat{\alpha}\rangle_L$  & \multirow{2}{*}{$\frac{1}{2}$} & \multirow{2}{*}{~$(\bf{4}, \bf{2}, 1)$} & \multirow{2}{*}{~$\mathbb{F}_L$} \\[0.5em]
&$~|\alpha\rangle_R \otimes {\overline{\psi}}^1_0{\overline{\psi}}^2_0{\overline{\psi}}^3_0{\overline{\psi}}^a_0|\hat{\alpha}\rangle_L$ &  &  &    \\[0.5em]
& $~|\alpha\rangle_R \otimes |\hat{\alpha}\rangle_L$  & \multirow{4}{*}{$\frac{1}{2}$} &\multirow{4}{*}{~$(\bf{4}, 1, \bf{2})$} & \multirow{4}{*}{~$\mathbb{F}_R$} \\[0.5em]
& $~|\alpha\rangle_R \otimes {\overline{\psi}}^4_0{\overline{\psi}}^5_0|\hat{\alpha}\rangle_L$ & & &\\[0.5em]
&$~|\alpha\rangle_R \otimes {\overline{\psi}}^i_0{\overline{\psi}}^j_0|\hat{\alpha}\rangle_L$ & & & \\[0.5em]
&$~|\alpha\rangle_R \otimes {\overline{\psi}}^i_0{\overline{\psi}}^j_0{\overline{\psi}}^4_0{\overline{\psi}}^5_0|\hat{\alpha}\rangle_L$ & & & \\[0.5em]
\hline
\multirow{2}{*}{$~\overline{V_1+V_2}~$} & $~|\alpha\rangle_R \otimes |\beta\rangle_L$ & $0$ & ~$(\bf{4}, \bf{2}, 1)$ & ~Exotic spinor $\mathbb{E}$ \\[0.5em]
&  $~|\alpha\rangle_R \otimes |\beta\rangle_L$ & $0$ & ~$(\bf{4}, 1, \bf{2})$ & ~Complex scalar $\mathbb{K}$~ \\[0.5em]
\hline
\end{tabular}
\caption{Chiral ($\mathbb{Z}_2$-untwisted) multiplets of the $\mathcal{N}=1$, $D=4$ Pati-Salam model 
that remain massless after the CDC. 
Here $i,j\in SU(4)$ and $a\in SU(2)_L\otimes SU(2)_R$. 
The $|\alpha\rangle_R$ represent right-moving Ramond ground states (space-time spinors), 
while $|\hat{\alpha}\rangle_L$ (respectively $|\beta\rangle_L$) 
represent the left-moving Ramond excitations that do 
not (respectively do) overlap with the Pati-Salam gauge group.
Again the multiplets are essentially the decomposition of the $\bf 16$ of $SO(10)$. 
The same decomposition applies for the two massless generations of the $b_3$- and $b_4$- 
 twisted-sector matter fields.
 \label{table:I}}
\end{table}
\begin{table}[H]
\bigskip
\centering
\setlength{\fboxsep}{0pt}
\begin{tabular}{|c|l|c|l|l|}
\hline
~Sector~ & ~~States removed by CDC~~ & ~Spin~ & ~$SU(4)\otimes SU(2)_L\otimes SU(2)_R$~ & ~Particle \\
\hline
& & & &\\
\multirow{2}{*}{${V_1+V_2}$}  & $~|\alpha\rangle'_R \otimes |\beta\rangle_L$ & $\frac{1}{2}$ & ~$(\bf{4}, \bf{2}, 1)$ & ~Spinor  $\tilde{\mathbb{E}}$ \\[0.5em]
& $~|\alpha\rangle'_R \otimes |\beta\rangle_L$ & $\frac{1}{2}$ & $~(\bf{4}, 1, \bf{2})$ & ~Spinor $\tilde{\mathbb{K}}$~~\\[0.5em]
\hline
\multirow{8}{*}{~$V_0+V_2$}~ & $~|\alpha\rangle'_R \otimes {\overline{\psi}}^i_0{\overline{\psi}}^a_0|\hat{\alpha}\rangle_L$  & \multirow{2}{*}{$0$} & \multirow{2}{*}{$~(\bf{4}, \bf{2}, 1)$} & \multirow{2}{*}{~$\tilde{\mathbb{F}}_L$} \\[0.5em]
&$~|\alpha\rangle'_R \otimes {\overline{\psi}}^1_0{\overline{\psi}}^2_0{\overline{\psi}}^3_0{\overline{\psi}}^a_0|\hat{\alpha}\rangle_L$ &  &  &    \\[0.5em]
&$~|\alpha\rangle'_R \otimes |\hat{\alpha}\rangle_L$  & \multirow{4}{*}{$0$} &\multirow{4}{*}{$~(\bf{4}, 1, \bf{2})$} & \multirow{4}{*}{~$\tilde{\mathbb{F}}_R$} \\[0.5em]
&$~|\alpha\rangle'_R \otimes {\overline{\psi}}^4_0{\overline{\psi}}^5_0|\hat{\alpha}\rangle_L$ & & &\\[0.5em]
&$~|\alpha\rangle'_R \otimes {\overline{\psi}}^i_0{\overline{\psi}}^j_0|\hat{\alpha}\rangle_L$ & & & \\[0.5em]
&$~|\alpha\rangle'_R \otimes {\overline{\psi}}^i_0{\overline{\psi}}^j_0{\overline{\psi}}^4_0{\overline{\psi}}^5_0|\hat{\alpha}\rangle_L$ & & & \\[0.5em]
\hline
\end{tabular}
\caption{Chiral ($\mathbb{Z}_2$-untwisted) multiplets 
of the $\mathcal{N}=1$, $D=4$ Pati-Salam model 
which are given masses $\frac{1}{2}\sqrt{ R_1^{-2}+R_2^{-2} }$ by the CDC. 
Here $i,j\in SU(4)$ while  $a\in SU(2)_L\otimes SU(2)_R$.  
The $|\alpha\rangle'_R$ represent right-moving Ramond ground states that are not space-time spinors.}
\label{table:J}  
\end{table}

\subsection{Cosmological constant}

\begin{figure*}[t!]
\begin{center}
  \epsfxsize 5.0 truein \epsfbox {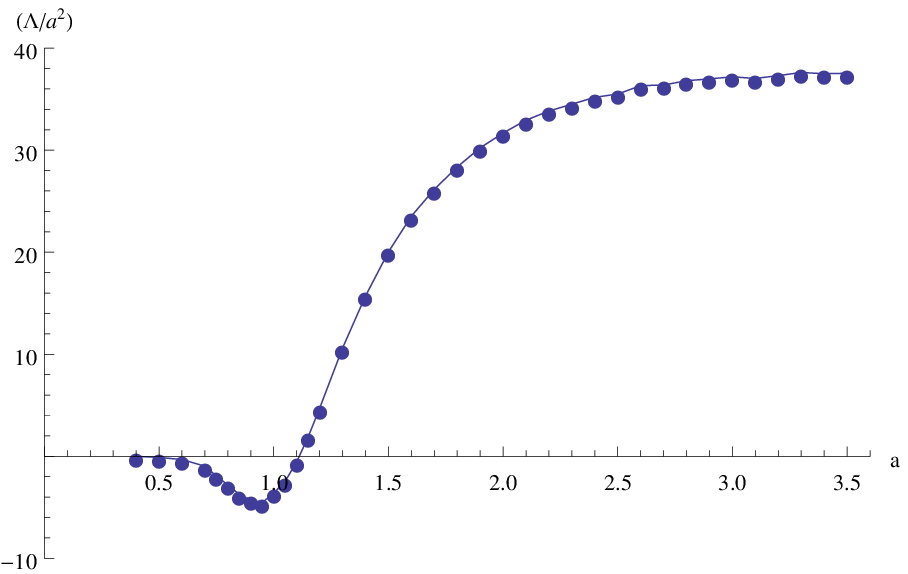}    
\end{center}
\caption{The rescaled cosmological constant $\Lambda/a^2$ for the Pati-Salam model versus $a\equiv \sqrt{\alpha'}/R$.
  For large $a$,  we find that $\Lambda/a^2$ tends to a constant indicating that the $a\to \infty$ limit of this model is 
  non-supersymmetric and tachyon-free.
  We also see that the entire curve is finite, which indicates that no tachyons emerge at any intermediate radii.
  Thus this model lacks Hagedorn-like instabilities.
  However we observe that the small-$a$ behaviour of this curve is radically different from the generic case.
  First, we see that $\Lambda$ does not have the usual Casimir $a^{4}$ behaviour, but rather is
  exponentially suppressed.
  Second, and somewhat surprisingly, 
  we observe that $\Lambda$ changes sign as $a\to 0$ increases past unity.
  Indeed, we see that the cosmological constant appears to have
  a stable minimum near (but not precisely at) the self-dual radius,
  and moreover that the cosmological constant crosses zero at yet another (slightly higher) radius.
  It is not clear whether there might exist enhanced symmetry at either of these specific radii.}
\label{fig:cdcinterplot}
\end{figure*}

 It is also interesting to examine the cosmological constant $\Lambda(a)$ of the Pati-Salam model as a function of the inverse radius $a=\sqrt{\alpha'}/R$.
Our results are shown in Fig.~\ref{fig:cdcinterplot}, and are consistent with the gross
features that one would expect from the above discussion:  the cosmological constant is
finite for all radii, exponentially suppressed in the large-radius limit, and radius-independent in the small-radius limit.  
This last observation suggests the existence of a zero-radius endpoint model (a.k.a. $M_2$) with an entirely non-supersymmetric but tachyon-free spectrum --- one which most likely corresponds to 
a 6D fermionic string constructed with discrete torsion.
More surprisingly, however,
just above (but not at) the self-dual radius, we find a stable anti-de Sitter minimum. 
This turn-over could indicate a restoration of gauge symmetry and/or supersymmetry, and is similar to the situation encountered in the Type~II models of Ref.~\cite{Angelantonj:2006ut}.

\subsection{Scalar masses}

Finally let us turn to the stringy threshold corrections that generate  scalar 
masses, {\it etc}\/.   These are all in principle calculable. Indeed, at 
one-loop order, their chief contribution at large radius 
can be understood by a field-theoretic calculation as it is dominated by the physical modes 
propagating in the loops. 
The string-theoretic calculation of these effects can be carried out in a fashion analogous to the usual gauge beta-function 
calculation --- namely by directly determining the two-point function for the scalar, but 
with the appropriate Scherk-Schwarz modified partition function. 
As one might expect, the result no longer vanishes, and the amplitude can be written as 
\begin{equation}
A(k,-k)~=~-\left(2\pi\right)^{4}\frac{g_{\rm YM}^{2}}{16\pi^{2}}\int_{\mathcal{F}}\frac{d^{2}\tau}{4\tau_{2}}\sum_{\alpha,\beta,\mathbf{\ell}}\left(\frac{Y^{2}}{g_{\rm YM}^{2}}-\frac{1}{4\pi\tau_{2}}\right)\frac{|\vec{\ell}|^{2}}{\tau_2^{2}}Z_{\mathbf{\ell,0}}Z\left[\begin{array}{c}
\alpha\\
\beta
\end{array}\right].
\end{equation}
We can split the contributions into those from massless physical states and those from massive ones. 
The term $({4\pi\tau_{2}})^{-1}$ will be proportional to the overall cosmological constant and is therefore exponentially suppressed. The contribution from the massless-sector terms to the canonically normalised 4D Higgs squared-masses is given by 
\begin{eqnarray}
   M_{H_{1}}^{2} & = & \frac{1}{16\pi^{2}}\int_{\frac{1}{\mu^{2}}\approx1}^{\infty}\frac{d\tau_{2}}{4\tau_{2}^{5}}\sum_{\mathbf{\ell}={\rm odd},i}Y^{2}\, (N_{fH}^{i}-N_{bH}^{i})\, |\vec{\ell}|^{2}\, e^{-\frac{\pi}{\tau_{2}}|\vec{\ell}|^{2}}e^{-\pi\tau_{2}\alpha'm_{i}^{2}}\nonumber \\
 & \approx & \frac{2}{\alpha'}\, \frac{Y^{2}}{16\pi^{2}}\, (N_{fH}^{0}-N_{bH}^{0})\, \frac{\pi^{2} a^6}{320}~,
\end{eqnarray}
where the sum is divided into mass-levels $m_i$.  By contrast, the contribution from the massive states is
given by
\begin{equation}
M_{H_{1}}^{2}~=~\frac{2}{\alpha'}\frac{Y^{2}}{16\pi^{2}}(N_{fH}^{i}-N_{bH}^{i})\sum_{\mbox{{\bf \ensuremath{\ell}}}={\rm odd}}|\vec{\ell}|^{-5/2}(\sqrt{\alpha'}m_{i})^{7/2}e^{-2\pi\sqrt{\alpha'}m_{i}|\vec{\ell}|}~.
\end{equation}
The first of these expressions does not necessarily vanish even if its analogue does for the cosmological constant, because the Higgs couples differently to the states that are projected out by the CDC.  Note, however, the interesting possibility of exponentially suppressed Higgs masses as well.

\subsection{Final comment}

To conclude I should comment on several interesting recent developments concerning the question of large-volume ``decompactification'' in Refs.~\cite{Caceres:1996is,Kiritsis:1996xd,Kiritsis:1998en,Antoniadis:2000vd,recentFaraggi}. The decompactification problem (i.e., reaching large volumes while avoiding large gauge couplings) has also been discussed in the past literature in a somewhat different guise~\cite{Dienes:1998vh}. 
In the context of Scherk-Schwarz breaking of supersymmetry, this is to a certain extent a dimensional transmutation of the hierarchy problem.  At first glance, achieving order-one couplings in a generic theory appears to require a fine-tuning of one-loop corrections against tree-level ones.  However, there are mechanisms to overcome this fine-tuning problem that will be presented in forthcoming work~\cite{future}, some of which utilize ideas in Ref.~\cite{Dienes:1998vh} and particularly in Ref.~\cite{Dienes:2002bg}.  Moreover, one of the strong benefits of the models constructed in Ref.~\cite{Abel:2015oxa}, which have been the focus of this talk, is that they naturally generate exponentially suppressed scales, so that extremely large volumes may not even be required.  This will be discussed further in Ref.~\cite{future}.  Clearly a primary objective would then be to extend exponential suppression to the Higgs mass itself. This will  be explored in Ref.~\cite{future2}.  

\section*{Acknowledgments}

The work of KRD was supported in part by the US Department of Energy under Grant DE-FG02-13ER-41976, and by the US National Science Foundation through its employee IR/D program.   The opinions and conclusions expressed herein are those of the authors and do not represent any funding agency. EM is in receipt of an STFC fellowship.

\end{document}